\documentclass[aps,prl,reprint,groupedaddress,showpacs]{revtex4-1}

\usepackage{bm}
\usepackage{graphicx}
\usepackage{color}

\begin{document}

\title{Coupling of Active Motion and Advection Shapes Intracellular Cargo Transport}



\author{Philipp Khuc Trong$^{1,2}$, Jochen Guck$^{2}$, and Raymond E. Goldstein$^{1}$}
\affiliation{$^{1}$Department of Applied Mathematics and Theoretical Physics, 
University of Cambridge, Wilberforce Road, Cambridge CB3 0WA, 
United Kingdom}
\affiliation{$^{2}$Department of Physics, University of Cambridge, J.J. Thomson Avenue, 
Cambridge CB3 0HE, United Kingdom}

\date{\today}

\begin{abstract}
\noindent
Intracellular cargo transport can arise from passive diffusion, active motor-driven transport
along cytoskeletal filament networks, and passive advection by fluid flows entrained by such 
motor/cargo motion.  Active and advective transport are thus intrinsically coupled as related, 
yet different representations of the same underlying network structure. A 
reaction-advection-diffusion system is used here to show that this coupling affects the transport and 
localization of a passive tracer in a confined geometry. For sufficiently low diffusion, 
cargo localization to a target zone is optimized either by low reaction kinetics and decoupling of 
bound and unbound states, or by a mostly disordered cytoskeletal network with only weak 
directional bias. These generic results may help to rationalize subtle features of cytoskeletal networks, for example as observed for microtubules in fly oocytes.
\end{abstract}

\pacs{87.16.Wd, 47.61.Ne, 47.63.Jd, 87.19.rh \, Copyright (2012) by the American Physical Society.}

\maketitle

\begin{figure}[b]
	\includegraphics[width=0.45\textwidth]{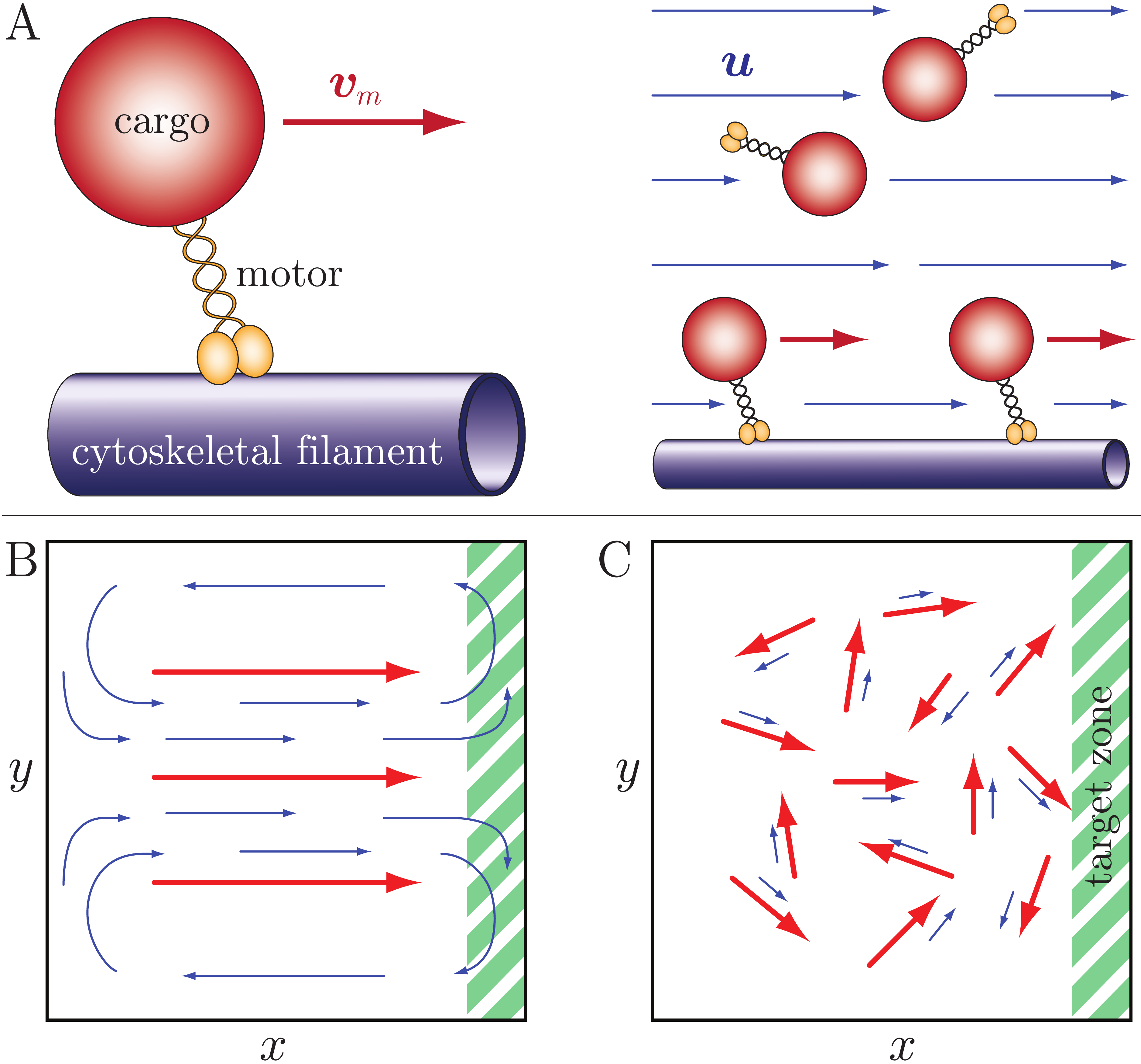}
	\caption{Coupling of active transport and advection and its system-level implications. A) Active motor-driven transport of cargo bound to a cytoskeletal filament (left) entrains surrounding fluid and causes advective transport of unbound cargo (right). B) A perfectly aligned cytoskeletal network (red arrows) causes recirculatory fluid flows (blue arrows) out of a target zone (green dashed area). C) A mostly disordered cytoskeletal network with only weak directional bias suppresses range and magnitude of fluid flows.}
	\label{fig:concept}
\end{figure}

Intracellular transport of proteins, vesicles or entire organelles is required by virtually all cells to 
perform functions as diverse as cell division, intracellular trafficking and patterning of 
morphogens during development. To realize these different functions, eukaryotic cells can utilize 
three different forms of cargo transport: passive diffusion by thermally driven Brownian 
motion, active transport by motor proteins on cytoskeletal networks \cite{Vale2003}, and 
passive advection by intracellular flows of bulk cytoplasm. 
Such cytoplasmic flows have been studied in plants \cite{LubiczGoldstein2010} as 
well as animals, including rats, mice, worms and flies
\cite{Bradke1997,Ajduk2011,Niwayama2011,Serbus2005}. 
While some cytoplasmic flows result from contractions of actin networks   
\cite{Ajduk2011,Mayer2010,Niwayama2011}, cytoplasmic streaming in flies, 
Characean algae and pollen tubes is driven by forces from the motion of the actively 
transported cargo itself \cite{Palacios2002,Shimmen2007} (Fig. \ref{fig:concept}A). Hence, active and advective 
transport can be intrinsically coupled as two related, yet different representations of the 
underlying cytoskeletal network. This raises intriguing questions of how changes in 
cytoskeletal network architecture and binding kinetics affect the distribution of cargo 
when active and advective transport are coupled (Fig. \ref{fig:concept}B,C).

Existing theoretical work has largely focused on the physical mechanisms of flows 
\cite{Nothnagel1982} and either on the combination of diffusion and active transport 
\cite{Dinh2006, Klumpp2005, Brangwynne2009a}, or on the combination of diffusion and 
advective transport \cite{Goldstein2008,vandeMeent2008}. The system-level implications of 
interactions between all three transport mechanisms are poorly understood \cite{Heaton2011}. 
Here, we study implications of coupled active and advective transport for cargo 
localization to a target zone in a confined geometry, a situation relevant to establishment 
and maintenance of cellular asymmetries.  Examples include asymmetric cell divisions, cellular 
morphogenesis, embryonic and pre-embryonic development \cite{Li2010, Ganguly2012}. 
A perfectly aligned cytoskeletal network may be optimal for cargo localization to a target zone if considered alone. Our main finding, however, is that a perfectly aligned network can become suboptimal for localization when coupled to its corresponding recirculatory fluid flow that washes away the cargo once it is dropped off in the target zone (Fig. \ref{fig:concept}B). Instead, a mostly disordered network with only weak directional bias can become optimal for persistent accumulation of cargo in the target zone by balancing an on-average directional active transport with the suppression of fluid flow caused by it (Fig. \ref{fig:concept}C).

To formalize this concept, we construct a reaction-advection-diffusion model for the transport of a passive scalar tracer that is advected by two coupled, yet different velocity fields. A motor-velocity field $\bm{v}_m$ that advects the bound-state cargo concentration $c_b$ captures active motion on a dense cytoskeletal network, while the fluid flow field $\bm{u}$ that advects the unbound cargo concentration $c_u$ represents the cytoplasmic flow. Cargo exchanges between bound and unbound states via interconversion reactions that conserve total mass. The partitioning of cargo between these two states is regulated by a parameter $0 \leq \beta \leq 1$. Together with a diffusion term in the unbound state, the nondimensional transport part of the model is defined as:
\begin{eqnarray}
\frac{\partial c_b}{\partial t} &+& \bm{\nabla} \cdot \left( \bm{v}_m \, c_b \right) = 
2\, Da \left[ - (1-\beta) \, c_b + \beta \, c_u \right] \label{eq:cb} \\
\frac{\partial c_u}{\partial t} &+&\bm{\nabla} 
\cdot \left( \bm{u} \, c_u \right) = 2\, Da \left[ (1-\beta) \, c_b -\beta \, c_u \right] + Pe^{-1} \bm{\nabla}^2 c_u . \nonumber
\end{eqnarray}
Here, the nondimensional motor P\'eclet number $Pe = VL/D$ and Damk\"ohler 
number $Da = LK/V$ are determined by the typical motor velocity $V$, mean reaction 
rate $K$, system length $L$ and diffusion constant $D$.  
The advection fields $\bm{v}_m$ and $\bm{u}$ are coupled since $\bm{u}$ is the solution to 
the Stokes equations for a viscous incompressible ($\bm{\nabla} \cdot \bm{u}=0$) Newtonian fluid 
driven by forces from the motor velocity field. Suitably rescaled these are
\begin{equation}
	0= -\bm{\nabla} p + \bm{\nabla}^2 \bm{u}  + \bm{f}~, \ \ \ \ \ 
\bm{f} = a \, \bm{v}_m . \label{eq:stokes1}
\end{equation}
 In general, the forces will depend on the concentration of bound cargo, with $a = a(c_b)$, but this more complex case is left to future work. Here we focus on the simplest case of constant proportionality between forces and motor velocities, and set $a = 1$ for convenience.
The solution of (\ref{eq:stokes1}) with no-slip conditions on the domain boundary was obtained with a finite volume discretization on staggered grids in Matlab using the SIMPLE algorithm \cite{Versteeg2007}. 
\begin{figure}
	\includegraphics[width=0.5\textwidth]{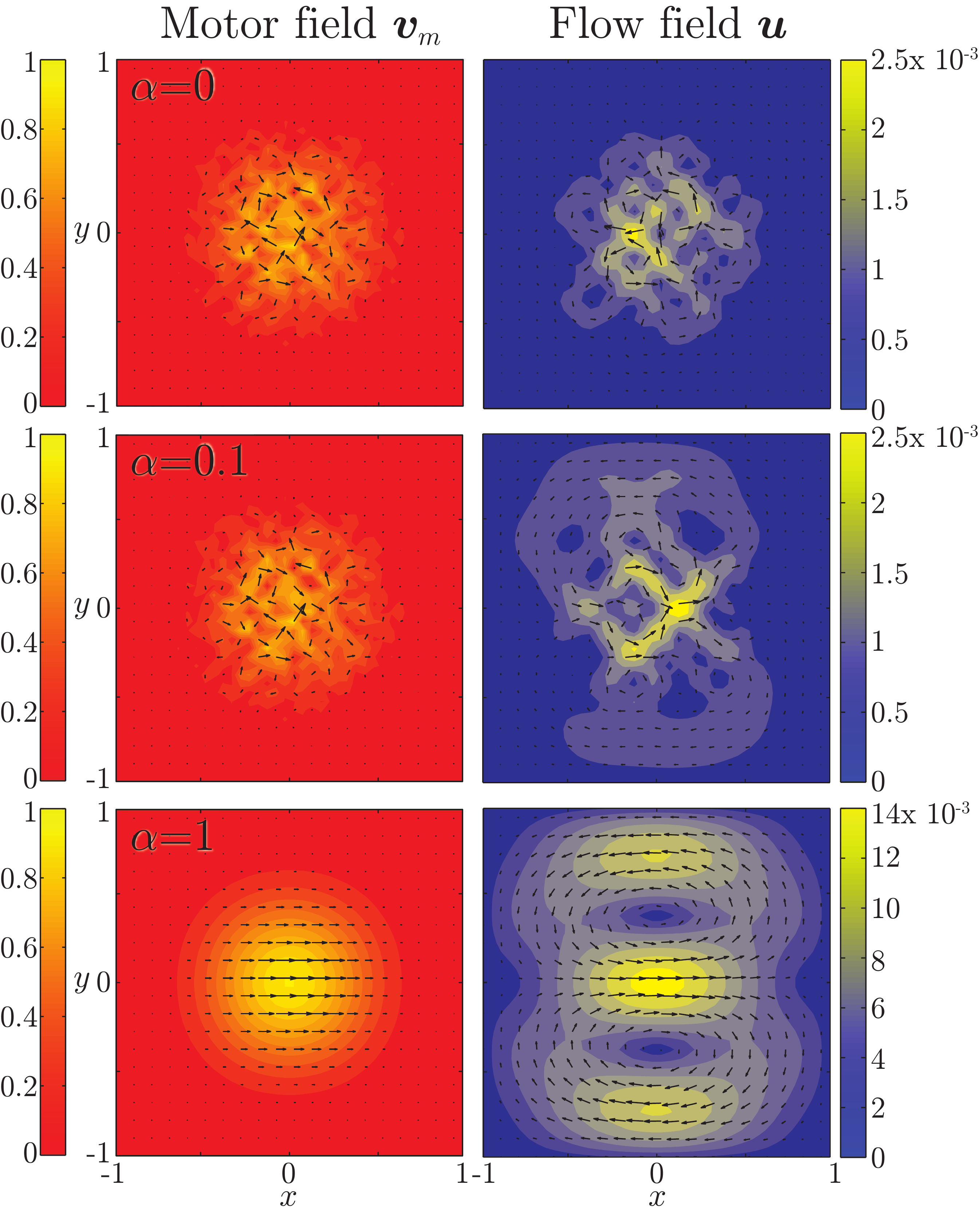}
	\caption{A perfectly aligned motor velocity field causes recirculatory fluid flow. Shown are topology (vector field) and magnitude (color coding) of motor velocity fields (left column) and the corresponding flow fields (right column) for varying network order parameter, $\alpha = 0$, $0.1$ and $1$, 
as indicated. To aid visibility only every second vector of the field is shown.}
	\label{fig:force_flow}
\end{figure}

Consider first the fluid flow field $\bm{u}$ for various degrees of order in the 
motor velocity field $\bm{v}_m$. Before normalizing to a peak magnitude of 1, we define 
on a two dimensional square $\bm{v}_m(x,y) = \bm{h}_1(x,y) \, h_2(x,y)$ 
wherein $h_2$ attenuates the magnitude of $h_1$ in the form
\begin{eqnarray*}
 4h_2(x,y) &=& \left\{{\rm erf}[m(b-x)] + {\rm erf}[m(b+x)] \right\} \\
	  & & \times \left\{{\rm erf}[(m(b-y)] + {\rm erf}[m(b+y)]\right\}  ~,
 \label{eq:h2}
\end{eqnarray*}
with ${\rm erf}(x)$ denoting the error-function, $m = 3$, and $b = 0.3$. 
The function $\bm{h}_1$ is a weighted sum of the form
\begin{eqnarray*}
 \bm{h}_1(x,y) = (1-\alpha) \left(\begin{array}{c} \sin(kx) \, \cos(ky) + 
\zeta_x \\ -\cos(kx) \, \sin(ky) + \zeta_y \end{array}\right) + 
\alpha \left(\begin{array}{c} 1 \\ 0 \end{array}\right),\nonumber
\end{eqnarray*}
where $k = 4\pi$ and $0 \leq \alpha \leq 1$ acts as an order parameter for the directional 
bias of the motor velocity field. For $\alpha = 0$, $h_1(x,y)$ consists of an array of 
vortices perturbed by random numbers $\zeta_{x,y}$ from the open interval $(-0.5,0.5)$ such 
that streamlines of neighboring vortices connect (Fig. \ref{fig:force_flow}, top left). Similar 
vortex arrays have been employed extensively for example in percolation theory 
\cite{Isichenko1992}. Using this as the force field input to the Stokes equations, we find a fluid flow 
field that mirrors the vortex structure of the forcing, but with a magnitude reduced by a 
factor of $10^3$ (Fig. \ref{fig:force_flow}, top right). 

For $\alpha = 1$, the motor field is perfectly aligned along the $x$-direction 
(Fig. \ref{fig:force_flow}, bottom left), giving rise to a Stokes flow field that 
in the center is aligned along the abscissa as well. In the periphery, however, mass-conservation 
and incompressibility result in pronounced recirculatory flows in the opposite direction 
(Fig. \ref{fig:force_flow}, bottom right).
This demonstrates that the topologies of the motor velocity and fluid flow fields can differ strongly.

For intermediate and even small values of $\alpha$ (Fig. \ref{fig:force_flow}, middle left), the 
averaging properties of Stokes flow still yield recirculatory flow fields 
similar to the perfectly aligned case (Fig. \ref{fig:force_flow}, middle right), albeit with ten-fold lower magnitudes. Thus, while the flow topology remains approximately constant over a wide 
range of the directional bias, variations of $\alpha$ represent a possible mechanism to tune the 
magnitude of the fluid speed and hence its impact on cargo transport. 
\begin{figure}
	\includegraphics[width=0.5\textwidth]
{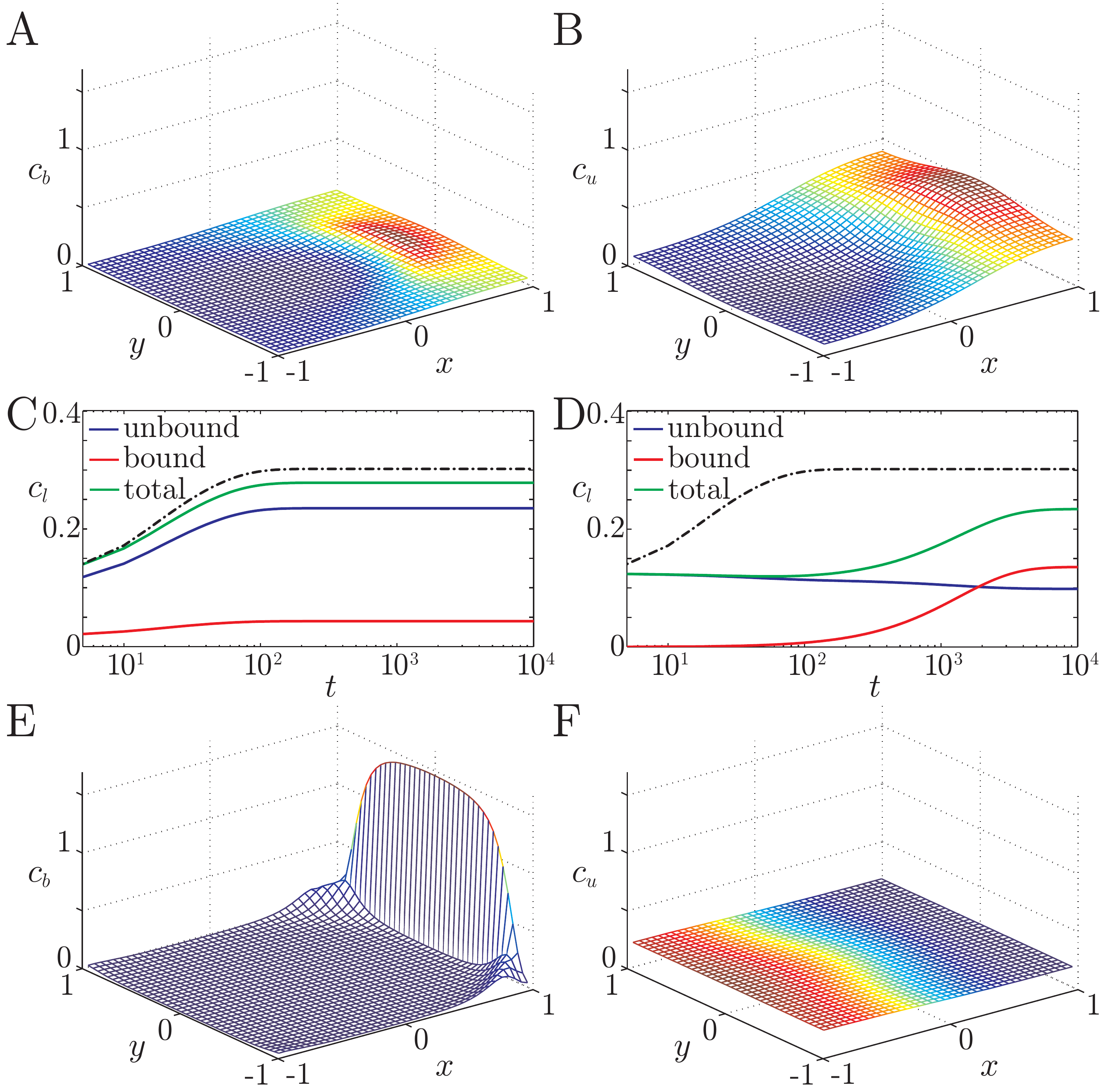}
	\caption{The parameter $Da$ regulates the coupling of bound and unbound states. 
For strong coupling $Da=1$, the steady-state distributions of bound cargo $c_b$ (A) and unbound cargo $c_u$ (B) are shown, with 
coloring on an arbitrary scale for each panel individually. 
(C) Fraction of cargo localized in the target zone $c_l$ in the simulation in A, B for the two-state 
system (\ref{eq:cb}) (solid lines) as well as for the effective one-state system 
(\ref{eq:c}) (dashed line). (D) Fraction of cargo localized in the target zone $c_l$ 
analogously to panel C, but for the simulations of bound cargo
(E) and unbound cargo (F) for weak coupling $Da = 4 \times 10^{-4}$ with coloring as in 
panels A-B. All transport simulations use $Pe = 10^2$, $\beta = 0.15$ and the motor velocity and fluid flow 
field with $\alpha = 1$ (Fig. \ref{fig:force_flow}, bottom row).}
	\label{fig:state_mixing}
\end{figure}

We next explore the consequences of these flow fields with fixed $\alpha$ for the localization of a chemical
species to a target zone. Depending on the system described, different initial conditions may be of
interest, including a homogeneous distribution or a deposit localized in a starting zone. Final concentration patterns are insensitive to this choice, and results are shown for the homogeneous case with cargo in the unbound state.
Cargo found at the end of a simulation in the stripe $ 0.75 \leq x \leq 1$ is considered as localized in the target zone (dashed area in Fig. \ref{fig:concept}B,C). We first study the effects of the Damk\"ohler number 
$Da$ that regulates the strength of chemical exchange between bound and unbound states. 

\begin{figure}
	\includegraphics[width=0.5\textwidth]{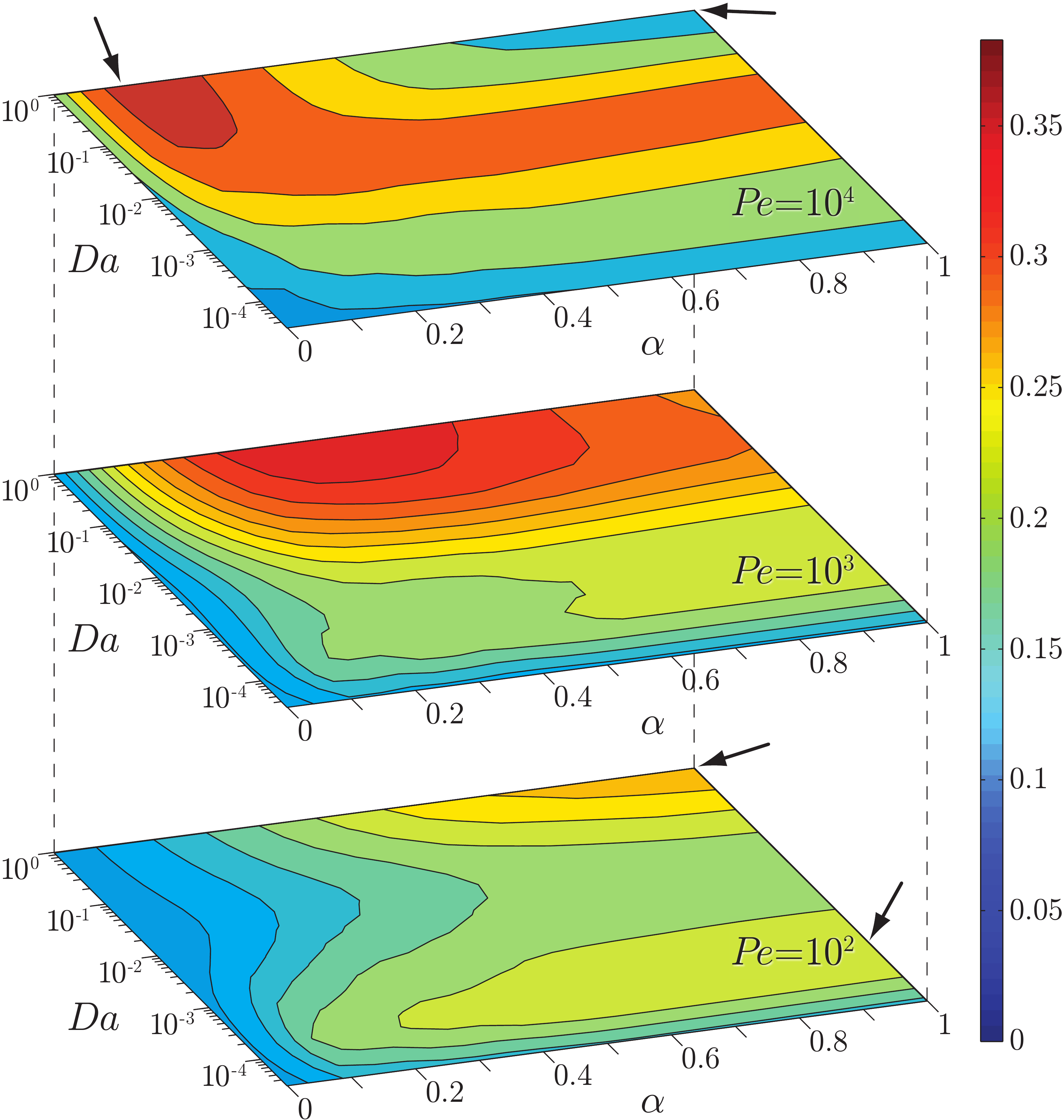}
	\caption{Parameter space for transport and localization on a 2D square. Contour plots show 
the fraction of total cargo localized in the target zone $0.75 \leq x \leq 1$ at the end of simulation 
time $t = 10^4$, with $\beta = 0.15$, as a function of the parameters $\alpha$ and $Da$ for three different motor 
P\'eclet numbers $Pe = 10^2$, $10^3$ and $10^4$ as indicated. Arrows highlight 
parameter values used in Figs. \ref{fig:state_mixing}A-C, 
\ref{fig:state_mixing}D-F, \ref{fig:suboptimal}A-B and \ref{fig:suboptimal}C-D.}
\label{fig:pspace_2D_square}
\end{figure}

When reactions are fast, cargo transport on a perfectly aligned motor network ($\alpha = 1$) and its corresponding flow field (Fig. \ref{fig:force_flow}, bottom row) 
show that the steady-state distributions in bound and unbound states are virtually identical, only scaled by the amounts of cargo in the respective states (Fig. \ref{fig:state_mixing} A, B). Cargo deposition in the 
target zone also occurs with the same dynamics for the two states (Fig. \ref{fig:state_mixing}C, solid lines). 
In the limit of very fast reactions ($Da\gg 1$) the system can be reduced to a single equation 
for the total cargo concentration $c = c_u + c_b$,
\begin{equation}
\frac{\partial c}{\partial t}
+ \bm{\nabla} \cdot \left[ \left( \beta \, \bm{v}_m + (1-\beta) \, \bm{u} \right) c \right] 
= (1-\beta)Pe^{-1}\nabla^2 c ~,
\label{eq:c}
\end{equation}
in which motor velocity and fluid flow fields mix to form an effective advection field supplemented by an effective diffusion  term \cite{Klumpp2005}.  This approximation works well even for 
$Da = 1$ (Fig. \ref{fig:state_mixing}C, dashed line).  

Transport simulations for slow reactions ($Da \ll 1$) show bound cargo accumulating at the extreme distal boundary, while unbound cargo remains mostly homogeneously distributed by diffusion (Fig. \ref{fig:state_mixing}E, F). Similarly, the 
dynamics of cargo accumulation separates into a roughly constant contribution from the unbound state, and into a slow increase due to the gradual recruitment of 
cargo to the bound state (Fig. \ref{fig:state_mixing}D). Hence, cargo transport in bound and 
unbound state proceeds virtually independently from one another. Thus, by regulating the strength 
of chemical reactions between bound and unbound states, $Da$ controls the degree of coupling 
of motor velocity and fluid flow fields.

We now vary both the network order parameter $\alpha$ and the coupling strength 
$Da$.
For $Pe = 10^2$ we find  (Fig. \ref{fig:pspace_2D_square}, bottom) that the highest amount of 
cargo localization occurs for a perfectly aligned motor field ($\alpha = 1$) and fast reaction 
kinetics ($Da = 1$). Strikingly, however, this combination of perfect alignment and strong mixing of 
bound and unbound states ceases to be the optimal configuration for cargo accumulation if $Pe$ is increased.

\begin{figure}
\includegraphics[width=0.5\textwidth]{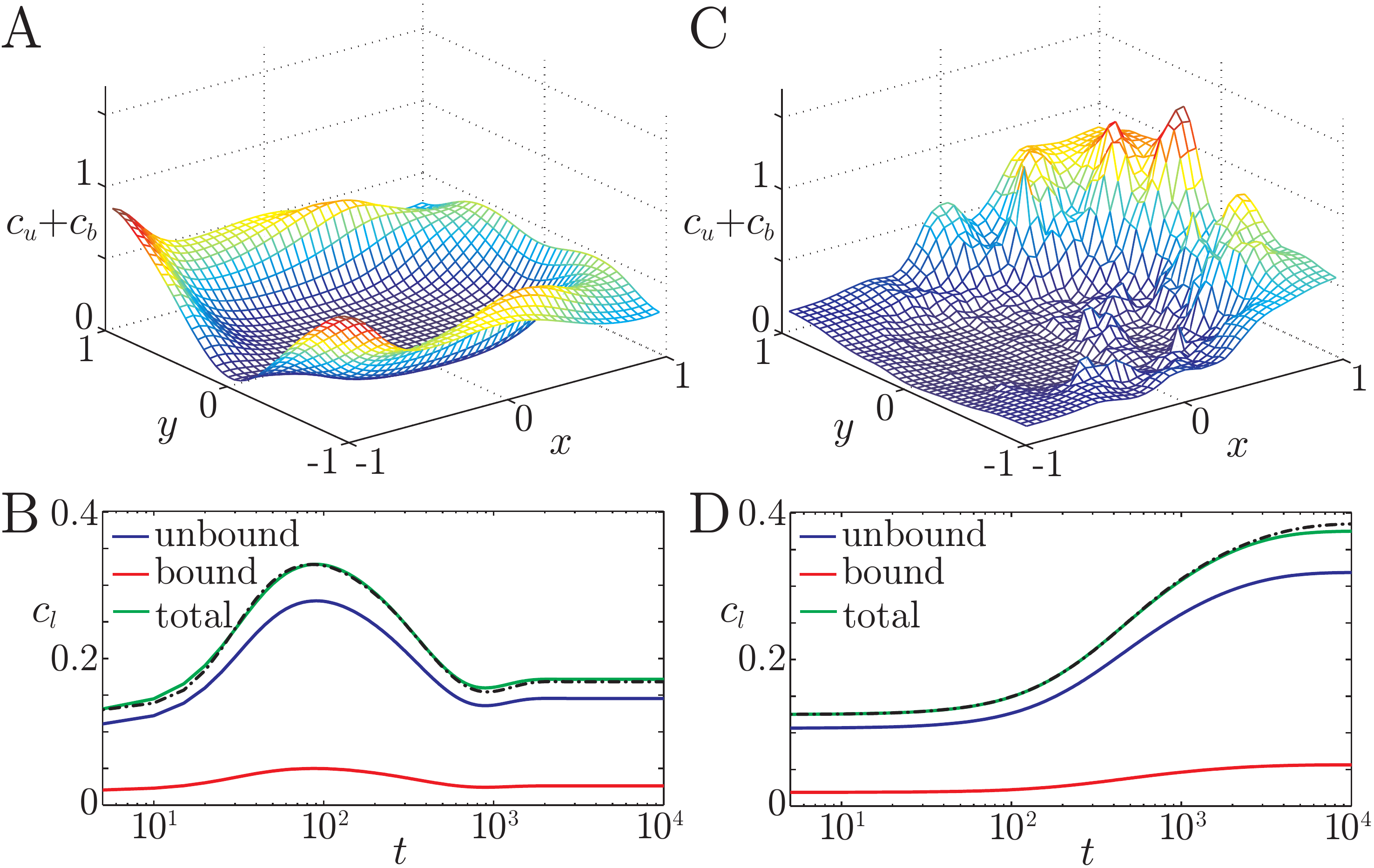}
\caption{A perfectly aligned motor velocity field is not optimal at high motor P\'eclet numbers. 
(A) Steady-state distribution of total cargo $c = c_u + c_b$ with coloring on an arbitrary scale for 
the motor velocity and fluid flow fields with $\alpha = 1$. 
(B) Fraction of localized cargo $c_l$ in the two-state system Eqs. (\ref{eq:cb}) (solid lines) as 
well as the effective one-state system Eq. (\ref{eq:c}) (dashed line) for the simulation in A. (C) Steady-state distribution 
of total cargo analogous to panel A, but for $\alpha = 0.1$.  (D) Fraction of localized cargo analogous to panel B, but for the simulation in C. All transport simulations use $Pe = 10^4$,
$Da = 1$ and $\beta = 0.15$.}
\label{fig:suboptimal}
\end{figure}

For values of $Pe = 10^3$ and $10^4$, respectively, the 
regime of high cargo accumulation in the target zone first moves towards smaller $\alpha$ 
(Fig \ref{fig:pspace_2D_square} middle), and finally  (Fig. \ref{fig:pspace_2D_square} top) 
forms a ridge circumventing the point 
($\alpha = 1$, $Da = 1$). Simulations at this point 
for $Pe = 10^4$ show a rapid accumulation of cargo at $t \sim 10^2$ 
(Fig. \ref{fig:suboptimal}B). This accumulation, however, remains transient due to the 
impact of the recirculatory backflows that move the bulk cargo towards the sides of the  
domain and eventually out of the target zone (Fig. \ref{fig:suboptimal}A). 
Strong accumulation of cargo in the target zone still occurs for lower reaction kinetics $Da\approx 10^{-2}$ (Fig. \ref{fig:pspace_2D_square}, top) that partially decouple bound and 
unbound states. Alternatively, high reaction kinetics combined with a strong reduction in directional bias to $\alpha \approx 0.1$ also lead to strong accumulation, albeit at the expense of slow dynamics (Fig. \ref{fig:suboptimal}C, D). Such changes in $\alpha$ have limited effects on the recirculatory flow pattern (Fig. \ref{fig:force_flow}). Instead, the reduction in fluid flow velocities stabilizes cargo accumulation 
in two ways: first by reducing directly the amount of material transported away from the target zone, 
and second by increasing the time for cargo to bind to the motor velocity field, hence increasing 
the amount of material that is returned to the target site. This counter-intuitive effect occurs over the wide range $0.1 \leq \beta \leq 0.75$ for which the fraction of localized cargo at low values of $\alpha$ is more than 10 percentage points higher than at $\alpha = 1$. The qualitative features of the parameter space also remain unchanged for simulations performed in a circular geometry, thereby highlighting the generality of the concept. 

Any biological cell that requires long-time or persistent cargo localization, for example prior to an asymmetric 
cell division, or to provide positional information during development, needs to limit dispersive 
effects. In general, biochemical mechanisms may contribute to stabilize cargo accumulation at the target site. Yet, the coupling between active and advective transport in our model indicates that an 
only weakly biased cytoskeletal 
network provides an alternative, physical strategy to balance an on-average directed active transport with suppressed cytoplasmic flows. 
Rough estimates for organelles or vesicles in Characean algae ($Pe \approx 5 \times 10^3$) or mRNA in fly oocytes ($Pe \approx 5 \times 10^3$) show that biological systems can reach the high P\'eclet number regimes explored here. 
This concept may therefore help to rationalize subtle directional biases recently discovered in microtubule networks of fly oocytes \cite{Parton2011}.


We thank H. Doerflinger, J. Dunkel, S. Ganguly, N. Giordano, I.M. Palacios, D. St. Johnston, and 
F.G. Woodhouse,  for discussions.  This work was supported in part by the 
Leverhulme Trust, the European Research
Council Advanced Investigator Grant 247333 (R.E.G.), the Boehringer Ingelheim Fonds (P.K.T.),
and the EPSRC.


\thebibliography{99}

\bibitem{Vale2003}  R. D. Vale, Cell {\bf 112}, 467 (2003).

\bibitem{LubiczGoldstein2010}  J. Verchot-Lubicz and R. E. Goldstein, Protoplasma {\bf 240},
99 (2010).

\bibitem{Bradke1997}  F. Bradke and C. G. Dotti, Neuron {\bf 19}, 1175 (1997).

\bibitem{Ajduk2011}  A. Ajduk, T. Ilozue, S. Windsor, Y. Yu, K. B. Seres,
R. J. Bomphrey, B. D. Tom, K. Swann, A. Thomas,
C. Graham, and M. Zernicka-Goetz, Nat Commun {\bf 2},
417 (2011).

\bibitem{Niwayama2011} R. Niwayama, K. Shinohara, and A. Kimura, Proc. Natl.
Acad. Sci. U.S.A. {\bf 108}, 11900 (2011).

\bibitem{Serbus2005} L. R. Serbus, B. J. Cha, W. E. Theurkauf, and W. M.
Saxton, Development {\bf 132}, 3743 (2005).

\bibitem{Mayer2010} M. Mayer, M. Depken, J. S. Bois, F. Julicher, and S. W.
Grill, Nature {\bf 467}, 617 (2010).

\bibitem{Palacios2002} I. M. Palacios and D. St Johnston, Development {\bf 129},
5473 (2002).

\bibitem{Shimmen2007} T. Shimmen, J. Plant Res.{\bf  120}, 31 (2007).

\bibitem{Nothnagel1982} E. A. Nothnagel and W. W. Webb, J. Cell Biol. {\bf 94}, 444
(1982).

\bibitem{Dinh2006} A. T. Dinh, C. Pangarkar, T. Theofanous, and S. Mi-
tragotri, Biophys. J. {\bf 90}, L67 (2006).

\bibitem{Klumpp2005} S. Klumpp and R. Lipowsky, Phys. Rev. Lett. {\bf 95}, 268102
(2005).

\bibitem{Brangwynne2009a} C. P. Brangwynne, G. H. Koenderink, F. C. MacKintosh,
and D. A. Weitz, Trends Cell Biol. 19, 423 (2009).

\bibitem{Goldstein2008} R. E. Goldstein, I. Tuval, and J. W. van de Meent, Proc.
Natl. Acad. Sci. U.S.A. 105, 3663 (2008).

\bibitem{vandeMeent2008} J. W. van de Meent, I. Tuval, and R. E. Goldstein, Phys.
Rev. Lett. 101, 178102 (2008).

\bibitem{Heaton2011} L. L. M. Heaton, E. Lopez, P. K. Maini, M. D. Fricker,
and N. S. Jones, arXiv:1105.1647v2 [q-bio.TO] (2011).

\bibitem{Li2010} R. Li and B. Bowerman, Cold Spring Harb. Perspect. Biol.
{\bf 2}, a003475 (2010).

\bibitem{Ganguly2012} S. Ganguly, L. S. Williams, M. I. Palacios, and R. E.
Goldstein, preprint (2012).

\bibitem{Versteeg2007} H. K. Versteeg and W. Malalasekera, An Introduction
to Computational Fluid Dynamics: The Finite Volume
Method (Pearson Education Limited, 2007).

\bibitem{Isichenko1992} M. B. Isichenko, Rev. Mod. Phys. {\bf 64}, 961 (1992).

\bibitem{Parton2011} R. M. Parton, R. S. Hamilton, G. Ball, L. Yang, C. F.
Cullen, W. Lu, H. Ohkura, and I. Davis, J. Cell Biol.
{\bf 194}, 121 (2011).
\end{document}